\documentclass{article}
\usepackage{amsmath}
\usepackage{titling}
\usepackage[fontsize=12pt]{fontsize}
\usepackage[a4paper, left=1in, right=1in, top=1in, bottom=1in]{geometry} 
\usepackage{amsmath, amsfonts, amssymb}  
\usepackage{graphicx}  
\usepackage{cite}  
\usepackage{siunitx}

\title{\textbf{The Proton Radius Puzzle and Discrepancies in Proton Structure Measurements}}
\author{
    Roland B. Lumpay$^{1*}$, Jade C. Jusoy$^{1}$, Ruel Apas$^{1}$, Eulogio Auxtero Jr.$^{1}$ \\
    $^{1}$Department of Physics, Caraga State University - Main Campus \\
    8600 Butuan City, Agusan del Norte, Philippines \\
    Email: \texttt{$^{1*}$roland.lumpay@carsu.edu.ph}
}
\date{January 2025}

\begin{document}

\vspace{-10mm}  
\maketitle  

\begin{abstract}
The proton radius puzzle remains a key challenge in modern physics, highlighting both the precision and limitations of current experimental and theoretical approaches. Recent studies, such as those by Xiong \textit{et al.} \cite{xiong2019small} and Bezginov \textit{et al.} \cite{bezginov2019measurement}, have consistently found a smaller proton radius of about 0.84 femtometers, in line with muonic hydrogen measurements \cite{pohl2010size}, but discrepancies with earlier electron-proton scattering and atomic hydrogen spectroscopy persist. These unresolved differences reveal systematic errors or limitations in existing theoretical frameworks, as pointed out by Arrington and Sick \cite{arrington2015evaluation}. The divergence between results from muonic and electronic hydrogen measurements remains unexplained, with contributions from both experimental uncertainties and theoretical gaps. To address these challenges, a coordinated approach is needed—focused on reducing systematic uncertainties through advanced experimental setups, like improved scattering experiments at Jefferson Lab \cite{xiong2019small}, and developing new theoretical models, such as those proposed by Alarcón \textit{et al.} \cite{alarcon2019proton} and Lin \textit{et al.} \cite{lin2022new}. Integrating experimental precision with theoretical rigor offers the best path toward resolving this enduring puzzle and providing deeper insights into the fundamental forces of nature.
\end{abstract}

\vspace{0.5cm}

\section{Introduction}
Understanding the proton's structure is fundamental to nuclear physics and our broader understanding of matter. As a critical component of atomic nuclei, the proton's properties, particularly its charge radius, are crucial for testing the predictions of the Standard Model and Quantum Electrodynamics (QED), the quantum theory describing electromagnetic interactions \cite{pohl2013muonic, carlson2015proton}. Precise measurements of the proton radius have historically been obtained from electron-proton scattering experiments and hydrogen spectroscopy \cite{mohr1999codata, sick2003rms, bernauer2010high,beyer2017rydberg}. However, recent high-precision measurements from muonic hydrogen spectroscopy, where a muon orbits the proton instead of an electron, have yielded a proton radius significantly smaller than previous estimates \cite{pohl2010size}. This discrepancy, now known as the “Proton Radius Puzzle,” suggests potential gaps in our understanding of proton structure and challenges our existing models of particle interactions. Resolving this puzzle has implications for QED, as well as for theories beyond the Standard Model, potentially hinting at new physics or requiring refined theoretical approaches to the fundamental forces governing matter \cite{carlson2015proton}.

Given the unresolved nature of the Proton Radius Puzzle and the advancements in experimental techniques, a comprehensive review of the current findings is both timely and necessary. Recent experiments, including high-precision electron scattering \cite{bernauer2010high, xiong2019small}, and new muonic hydrogen spectroscopy data \cite{pohl2010size}, continue to produce varied results, challenging previous assumptions about proton structure \cite{mohr1999codata}. Despite numerous studies, no consensus has been reached, and the inconsistency persists, raising questions about the underlying physics or potential experimental limitations. Reviewing these developments now could clarify the current status of the puzzle, identify areas where theory and experiment diverge, and possibly guide future experimental and theoretical approaches. As the field progresses, understanding the proton radius has become a critical touchstone for both validating established physics and probing for new phenomena, making this review essential for guiding upcoming research directions in particle and nuclear physics.

\subsection{ Background of the Study}
The proton, identified over a century ago, remains a foundational component of visible matter \cite{rutherford2010collision}. It is a stable particle and one of the fundamental building blocks in atomic nuclei \cite{yang2024discovery}. Despite being stable, it has been extensively studied for decades \cite{carlson2015proton, karr2019progress, bernauer2020proton}. One of its properties, the charge radius, remains a puzzle \cite{carlson2015proton, pohl2010size, xiong2019small}. This puzzle arises because of the various measurements of the proton radius. The puzzle was further intensified in 2010 \cite{pohl2010size}, when a new measurement of the proton radius, obtained through muonic hydrogen spectroscopy, diverged from the previously accepted value based on electron-proton scattering and atomic hydrogen spectroscopy. This unexpected result sparked widespread interest in the research community, leading to questions about the validity of the long-established theoretical framework, Quantum Electrodynamics (QED) — the model that describes the interaction between light and matter \cite{pohl2010size, antognini2013proton}. The discrepancy in the proton radius raises doubts about the limits of QED's applicability, especially in systems involving different leptons, such as electrons and muons \cite{carlson2015proton}. If a smaller radius is consistently observed in muonic hydrogen compared to electron-proton scattering, it could suggest that QED may require revision \cite{antognini2013proton}. This would be necessary to address the methods already within its scope and account for potential unknown effects and higher-order corrections, particularly in muonic systems.

Historically, the size of a proton was determined by electron-proton (ep) scattering experiments and hydrogen spectroscopy. The electron-proton scattering technique, which has been used since the 1950s, involves firing high-energy electrons at a proton target and analyzing the angular distribution of the scattered electrons \cite{hofstadter1956electron}. This technique provides direct information about the electric and magnetic form factors of the proton, which are used to infer its charge distribution and size \cite{bernauer2010high, bernauer2014electric}. This method has been refined over the years, with increasingly precise measurements consistently supporting the larger proton radius of approximately 0.88 fm \cite{sick2003rms, bernauer2010high, zhan2011high}. Similarly, traditional hydrogen spectroscopy, which measures the energy levels of electrons in hydrogen atoms, has also supported and accepeted this larger radius \cite{mohr1999codata, sick2003rms,beyer2017rydberg}, for decades until 2010, when a groundbreaking experiment using muonic hydrogen spectroscopy—a technique involving a muon orbiting a proton—revealed a proton radius of about approximately 0.84 fm \cite{pohl2010size}, significantly smaller than the previously accepted value.

The persistence of this discrepancy, known as the proton radius puzzle, has driven extensive experimental and theoretical efforts to resolve the issue. Just a recent experiment by Xiong \textit{et al.} \cite{xiong2019small}, using electron-proton (ep) scattering, further refined the previous scattering results. While the findings are promising as they tend to agree with results from muonic hydrogen spectroscopy, the puzzle remains unsolved, necessitating continued investigations and the development of new experimental approaches to reconcile the discrepancy.

\subsection{Scope and Purpose}
This review offers a thorough analysis of the experimental methods and theoretical approaches involved in determining the proton radius, with key goals of evaluating major advancements, investigating possible solutions to the proton radius puzzle, and considering the wider implications for nuclear physics and foundational theories. Specifically, we address the following research questions: What are the main experimental approaches and theoretical perspectives have contributed to the current understanding of the proton radius? How might the discrepancies in measurements be reconciled? What are the implications of these findings for QED, the Standard Model, and potential new physics?

This review covers the history of measurement techniques, recent electron-proton scattering advances, and insights from muonic hydrogen spectroscopy. We also discuss various theoretical interpretations that have been proposed to account for discrepancies in measurements. However, we do not delve into highly specialized subfields, such as lattice QCD calculations, as they fall beyond the scope of this overview. Our methodology for literature selection includes focusing on seminal and high-impact studies from peer-reviewed journals, prioritizing works that contribute directly to understanding the proton radius through experimental and theoretical lenses.

This review is pertinent to nuclear and particle physics, as well as to related fields such as atomic physics and precision measurement science. The proton's size is essential not only for nuclear physics but also as a critical benchmark for assessing the validity of QED and the Standard Model, potentially revealing new insights into unexplored aspects of fundamental forces.

\section{Techniques for Proton Radius Extraction}
Accurately determining the proton's charge radius has historically relied on two primary experimental approaches: electron-proton scattering and hydrogen spectroscopy. However, the 2010 introduction of muonic hydrogen spectroscopy dramatically shifted the landscape, providing a more precise measurement but also exposing a critical discrepancy that has since become known as the proton radius puzzle.

\subsection{Atomic Hydrogen Spectroscopy}
For more than five decades, the study of atomic hydrogen spectroscopy, supported by its precisely calculable atomic structure, has been instrumental in the development and validation of quantum electrodynamics (QED), leading to highly accurate measurements of the Rydberg constant and the proton charge radius \cite{biraben2009spectroscopy}. Atomic hydrogen spectroscopy studies the energy levels and spectral lines of hydrogen atoms \cite{bransden2003physics}. It involves measuring the wavelengths of light emitted or absorbed when electrons transition between different energy levels in the hydrogen atom \cite{woodgate1970elementary}. This technique is crucial for understanding the fundamental properties of hydrogen, such as the Rydberg constant, and for determining the proton radius \cite{baklanov1974precise}. 

Building on these foundational principles, recent studies, such as the one by Fleurbaey \textit{et al.} \cite{fleurbaey2018new}, have advanced our understanding by employing more sophisticated techniques, such as continuous-wave (CW) laser spectroscopy at 205 nm, which is used to excite hydrogen atoms via the 1S--3S two-photon transition.
The CW laser was generated through sum frequency generation (SFG) in a $\beta$-barium borate crystal using a tunable titanium:sapphire (Ti:Sa) laser at 894 nm and frequency-doubled 532 nm radiation. The laser was stabilized using a series of Fabry--Perot cavities and a rubidium reference standard, with frequency measurements performed using a MenloSystems femtosecond frequency comb linked to cesium fountain standards. The hydrogen atoms were provided by an effusive beam from dissociated H$_2$ molecules, with fluorescence from the 3S--2P Balmer-$\alpha$ transition detected using a photomultiplier tube. Systematic corrections included second-order Doppler (SOD) effects, characterized via velocity distribution modeling, and light and pressure shifts, corrected by extrapolating data recorded at varying pressures and laser powers. Data analysis involved fitting the observed fluorescence signals to theoretical profiles accounting for velocity distribution parameters and systematic effects. Two separate datasets were collected (2013 and 2016--2017), analyzed independently, and combined to yield a transition frequency with a relative uncertainty of $9 \times 10^{-13}$. The proton charge radius derived from this measurement, \(r_p = 0.877(13)\) fm, aligns closely with the CODATA-recommended value \cite{mohr1999codata}. 
The experimental set-up is shown in Figure 1.

\begin{figure}[h!] 
\vskip1mm
\centering 
\includegraphics[width=0.8\columnwidth]{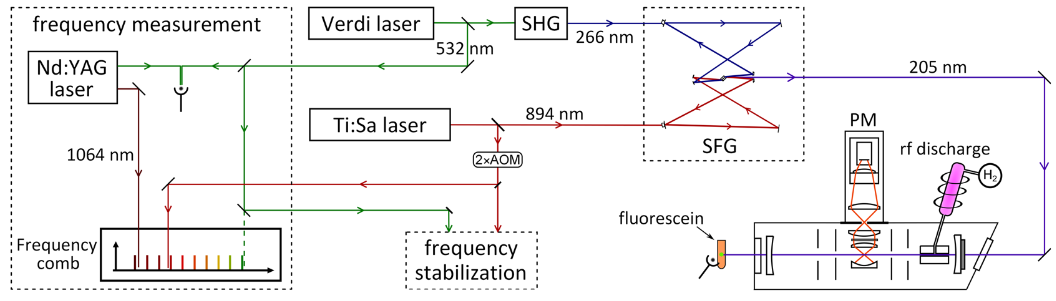} 
\vskip-3mm
\caption{The system stabilizes laser frequencies using Fabry-Perot cavities and a rubidium-stabilized reference laser. Since 2016, the Verdi laser's frequency has been measured via an Nd:YAG transfer laser rather than directly interfacing with the frequency comb. Key elements include second harmonic generation (SHG), sum frequency generation (SFG), an acousto-optic modulator (AOM), and a photomultiplier (PM) \cite{fleurbaey2018new}.}
\label{fig:image_label}
\end{figure}

Additionally, studies such as Beyer \textit{et al.} \cite{beyer2017rydberg} conducted precision spectroscopy of 2S--nP transitions in atomic hydrogen to refine the proton charge radius ($r_p$) and Rydberg constant ($R_\infty$). A cryogenic beam of metastable hydrogen atoms was used, significantly reducing Doppler effects and systematic uncertainties. The 2S state was populated using Doppler-free two-photon absorption, while transitions to higher excited states were driven by stabilized laser systems with counter-propagating beams to suppress first-order Doppler shifts. Lyman-$\gamma$ photons emitted during decay were detected using a high-efficiency, large solid-angle detector, while time-of-flight resolved methods isolated velocity classes. Systematic effects such as ac Stark and Zeeman shifts, as well as quantum interference in spontaneous emission, were rigorously modeled using optical Bloch equations and polarization-dependent studies. Further, transitions to higher excited states, like 2S--6P, were used to quantify dc Stark effects, with measurements extrapolated to zero velocity to minimize uncertainties. This methodology achieved reproducibility within a few parts in $10^{12}$, extracting $r_p = 0.8764(89)$~fm. The set-up of the experiment is shown in Figure 2.

\begin{figure}[htbp]
    \centering
    \includegraphics[width=0.8\columnwidth]{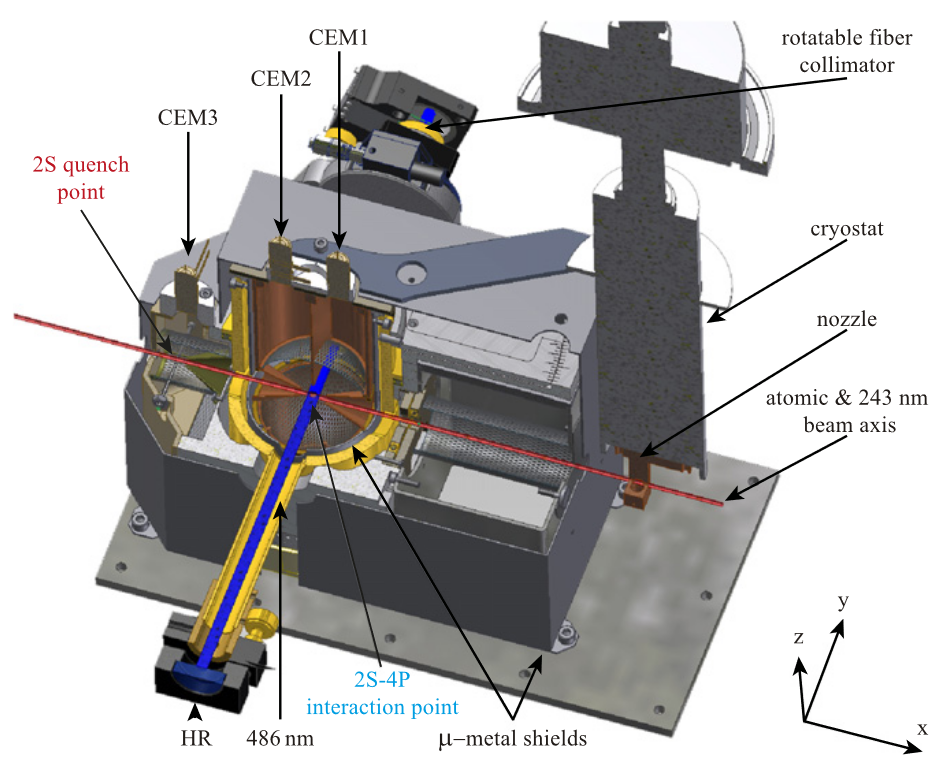} 
    \caption{Cryogenic $2S$ beam apparatus with large solid-angle detector setup: Hydrogen atoms are thermalized at $5.8 K$ inside a copper nozzle and collimated into a beam through two diaphragms. The $1S–2S$ transition is excited along the atomic beam axis, while $2S–4P$ excitation occurs at a near-perpendicular laser-to-beam angle. Lyman-$\gamma$ photons from $4P–1S$ decay are detected by channel electron multipliers (CEM1 and CEM2) via photoemission from graphite-coated regions. A new detector (CEM3) measures $2S$ atoms after de-excitation by a static electric field, normalizing flux and reducing noise. Polarization of the $486 nm$ excitation light is controlled by a rotatable collimation system \cite{beyer2017rydberg}.}
    \label{fig:image_label}
\end{figure}

Similarly addressing the proton radius puzzle, Thomas \textit{et al.} \cite{thomas2018high} conducted high-resolution spectroscopy of the $1S-3S$ transition in atomic hydrogen but employed a continuous-wave laser for their measurements. Using a counter-propagating photon configuration and thorough corrections for systematic effects, their approach achieved a fivefold improvement in precision, reducing the transition frequency uncertainty to 2.6~kHz. Their findings yielded a proton charge radius value consistent with the CODATA recommendation \cite{mohr2005codata}, further advancing the accuracy of hydrogen spectroscopy and shedding light on fundamental physical inconsistencies.

The precise nature of hydrogen spectroscopy, including highly accurate measurements of energy levels influenced by the proton’s charge distribution, allowed researchers to test the precision of these measurements, providing new insights into the proton radius \cite{grinin2020two}. Moreover, a study conducted by Bezginov \textit{et al.} \cite{bezginov2019measurement} utilized the FOSOF (frequency-offset separated oscillatory field) technique to determine the proton radius through a precise evaluation of the Lamb shift in atomic hydrogen. A beam of 55-keV protons was passed through molecular hydrogen gas, generating metastable hydrogen atoms in the $2S_{1/2}$ state. Charged particles were removed using deflector plates, and the neutral atoms traversed two RF regions tuned to the $2S_{1/2} \to 2P_{1/2}$ transition. The RF fields were offset in frequency, enabling phase-sensitive measurements. Surviving $2S_{1/2}$ atoms were mixed with $2P_{1/2}$ states via an electric field, emitting Lyman-$\alpha$ photons for detection.

The FOSOF method allowed the extraction of the transition frequency by identifying the phase shift at zero crossing, ensuring systematic errors were minimized through apparatus rotation and frequency alternation. Corrections for AC Stark shifts, time dilation, and beam velocity variations were applied using density-matrix modeling and extensive parameter sweeps. With 116 datasets averaged, the Lamb shift was measured with an uncertainty of $\pm3.2$ kHz, leading to a proton radius determination of $r_p = 0.833 \pm 0.010$ fm, consistent with muonic hydrogen measurements. The experimental set-up is shown in Figure 3.

\begin{figure}[htbp]
    \centering
    \includegraphics[width=0.8\columnwidth]{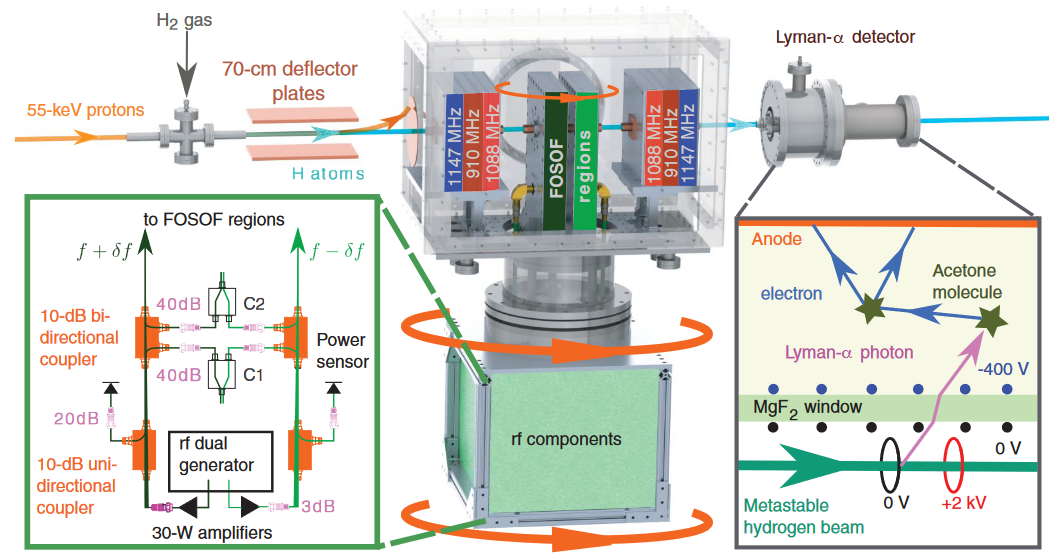} 
    \caption{Protons collide with \( \mathrm{H}_2 \) molecules to produce metastable \( 2S_{1/2} \) atoms, with deflector plates removing remaining protons. RF cavities (red and blue) filter out \( 2S_{1/2} \) (\( F=1 \)) atoms, leaving \( 2S_{1/2} \) (\( F=0 \)) atoms, which are excited to \( 2P_{1/2} \) (\( F=1 \)) in offset-frequency FOSOF regions (green). Surviving \( 2S_{1/2} \) (\( F=0 \)) atoms are mixed in an electric field, producing Lyman-\( \alpha \) photons detected via gas-ionization detectors. Rotating the entire FOSOF system ensures atoms experience reversed field orders, enhancing measurement reliability. Additional 910-MHz cavities (brown) check for systematic effects, with RF phase monitored by combiners C1 and C2 \cite{bezginov2019measurement}.}
    \label{fig:image_label}
\end{figure}

Another study by Grinin \textit{et al.} \cite{grinin2020two} utilized two-photon frequency comb spectroscopy to measure the 1S–3S transition in atomic hydrogen. The setup involved a cryogenically cooled atomic beam to minimize Doppler-related effects, with hydrogen atoms produced via dissociation of \( \text{H}_2 \) in a radio-frequency discharge. A frequency comb, stabilized to high precision, was employed to excite the atoms using UV light at 205 nm, generated through nonlinear frequency conversion of femtosecond laser pulses. The frequency stability was ensured by referencing the comb to a cesium atomic clock through an optical frequency chain. Fluorescence from the 3S–2P transition was detected using a photomultiplier tube. Systematic effects were rigorously controlled and minimized, including AC-Stark, Doppler, and pressure shifts, using direct measurement and extrapolation techniques. Data analysis included fitting spectral lines with theoretical profiles and performing global fits across datasets. This experimental approach achieved high resolution and accuracy, with systematic uncertainties reduced to below the \(7 \, \text{kHz}\) discrepancy characteristic of the Proton Radius Puzzle. The measurement precision was further enhanced by improvements in laser systems and detection efficiency, enabling the determination of transition frequencies with an uncertainty of \(0.72 \, \text{kHz}\). The study determined the centroid frequency of the 1S–3S transition as \( f_{1S-3S} = 2,922,743,278,665.79(72) \, \text{kHz} \), allowing improved determinations of the Rydberg constant (\(R_\infty = 10,973,731.568226(38) \, \text{m}^{-1}\)) and the proton charge radius (\(r_p = 0.8482(38) \, \text{fm}\)). The findings are consistent with the muonic hydrogen measurements but disagree with the CODATA 2014 values by \(2.9\sigma\). These results contribute to resolving systematic discrepancies in the Proton Radius Puzzle while setting a benchmark for precision spectroscopy and highlighting the potential for further advancements in experimental techniques. The set-up of the experiment is shown in Figure 4.

\begin{figure}[htbp]
    \centering
    \includegraphics[width=0.7\columnwidth]{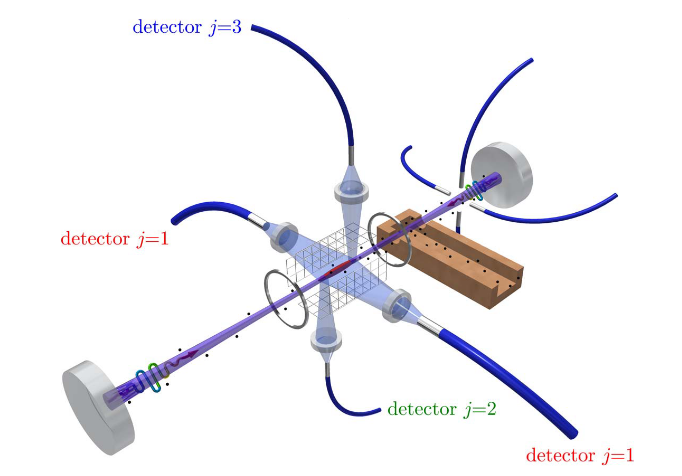} 
    \caption{The pulse collision volume (PCV), shaped like an ellipsoid with semi-axes \( w_0 = 80 \, \text{mm} \) and \( cT_{1/2} = 60 \, \text{mm} \), is enclosed in a Faraday cage with a highly transmissive mesh and two end-cap electrodes. To minimize stray electric fields, the quadratic DC-Stark shift is used by applying voltages along three directions and identifying minima in line shifts. Fluorescence from the PCV is captured by four lenses and directed through multimode fibers to single-photon counting modules, including one main detector and two auxiliary detectors. This setup allows for the interpolation of chirp-induced residual first-order Doppler shifts (CIFODS). Additionally, Doppler-broadened signals collected by 1 mm diameter fibers near the nozzle, which are laser-frequency independent, serve for normalization. By scaling both Doppler-free and Doppler-broadened signals with laser power, normalized line amplitudes offer a reliable measure of atomic flux, mitigating fluctuations in laser power and atom number flux \cite{grinin2020two}.}
    \label{fig:image_label}
\end{figure}

\subsection{Muonic-Hydrogen Spectroscopy}
Muonic hydrogen spectroscopy is a precise experimental method that uses a form of hydrogen where the electron is replaced by a muon—a particle 200 times heavier than the electron. This heavier mass causes the muonic hydrogen atom's Bohr radius to shrink significantly, increasing the overlap between the muon and the proton's wavefunctions. This results in an amplified sensitivity to the proton's internal structure and allows for highly accurate measurements of its charge radius and other nuclear properties \cite{pohl2010size}. The technique is used to measure properties like the proton charge radius, leading to significant insights and discussions in particle physics, particularly regarding discrepancies with traditional hydrogen measurements.

Two landmark studies have shaped the field of muonic hydrogen spectroscopy, with Pohl \textit{et al.}’s 2010 \cite{pohl2010size} research paving the way. In their study, Pohl \textit{et al.} employed laser spectroscopy to measure the Lamb shift in muonic hydrogen with unprecedented precision, yielding a proton charge radius of $r_p = 0.84184(67) \, \text{fm}$. This value, which is 10 times more precise than earlier measurements, deviates by \(5\sigma\) from the CODATA value of $r_p = 0.8768(69) \, \text{fm}$ \cite{mohr2008codata}. The experiment leveraged the smaller Bohr radius in muonic hydrogen to enhance sensitivity to the proton's finite-size effects, challenging the consistency of quantum electrodynamics (QED) calculations and suggesting potential refinements to the Rydberg constant or theoretical models. This discrepancy arises because the muonic hydrogen experiment exploits the muon's 200-fold greater mass compared to the electron, leading to a Bohr radius approximately 200 times smaller. The finite-size effects of the proton on the atomic energy levels, particularly the Lamb shift, are correspondingly enhanced, enabling a more sensitive determination of $r_p$. The energy difference of the Lamb shift ($\Delta \tilde{E}$) in muonic hydrogen is modeled as:

\begin{equation}
    \Delta \tilde{E} = 209.9779(49) - 5.2262 \, r_p^2 + 0.0347 \, r_p^3 \,\text{meV}
\end{equation}

where $r_p$ is in femtometers. This precise measurement of the Lamb shift was achieved through pulsed laser spectroscopy, which determined the transition frequency to be $49,881.88(76) \, \text{GHz}.$
The observed discrepancy in $r_p$ has profound implications, suggesting either a need to revise quantum electrodynamics (QED) calculations or the Rydberg constant, or the presence of new physics. The puzzle remains unresolved, challenging our understanding of proton structure and fundamental constants.Experimental Set-up for Laser System is shown in Figure 5.

\begin{figure}[htbp]
    \centering
    \includegraphics[width=0.7\columnwidth]{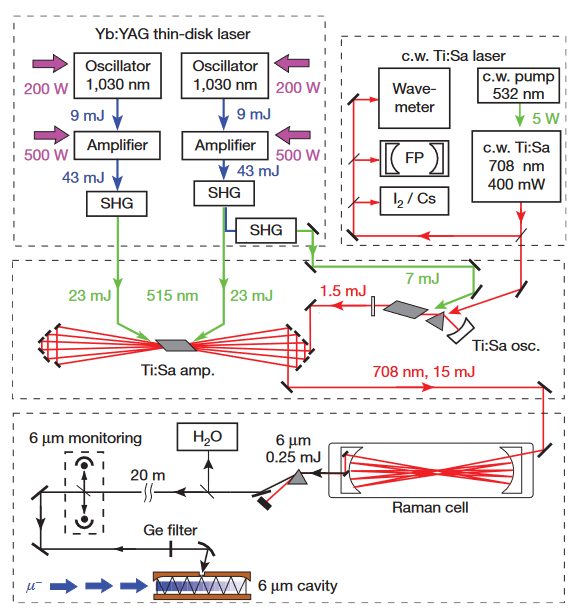}
    \caption{The continuous-wave (c.w.) light from the Ti:sapphire (Ti:Sa) ring laser (top right) seeds the pulsed Ti:Sa oscillator (middle). A detected muon triggers the Yb:YAG thin-disk lasers (top left), which, after second harmonic generation (SHG), pump the pulsed Ti:Sa oscillator and amplifier (middle). This setup produces 5 ns pulses at the wavelength defined by the c.w. Ti:Sa laser. These pulses are shifted to the desired wavelength (\(\lambda < 6 \, \text{mm}\)) via three sequential Stokes shifts in the Raman cell (bottom). The c.w. Ti:Sa laser is stabilized to an I\textsubscript{2}/Cs-calibrated Fabry-Perot (FP) reference cavity, with frequency calibration at \(\lambda = 6 \, \text{mm}\) using H\textsubscript{2}O absorption. \cite{pohl2010size}}
    \label{fig:proton_radius}
\end{figure}

The findings significantly contributed to the proton radius puzzle and underscored the need for further investigations~\cite{pohl2010size}. Building on these findings, Antognini \textit{et al.} \cite{antognini2013proton}  conducted precise measurements of 2S-2P transition frequencies in muonic hydrogen using an advanced experimental setup. The process began by stopping negative muons in H$_2$ gas at \SI{1}{hPa} and \SI{20}{\degreeCelsius}, forming highly excited muonic hydrogen atoms with approximately 1\% populating a long-lived 2S state. The experiment utilized a specialized laser system delivering \SI{5}{ns} pulses with wavelengths between $ 5.5 \,\mu m $ to $ 6 \,\mu m $, timed to illuminate the target gas volume $ 0.9 \,\,ms $ after muon entry. 

The study focused on measuring two key transitions: the singlet state, $\nu_s: \nu(2\mathrm{S}_{1/2}^{F=0} - 2\mathrm{P}_{3/2}^{F=1})$, and the triplet state, $\nu_t: \nu(2\mathrm{S}_{1/2}^{F=1} - 2\mathrm{P}_{3/2}^{F=2})$. Data collection involved adjusting laser frequencies every few hours with up to \SI{13}{hour} accumulation periods per frequency. Resonance detection was achieved by measuring \SI{1.9}{keV} x-rays coincident with laser pulses within a specific time window.

Rigorous calibration and error control were implemented, accounting for various systematic uncertainties including Zeeman shift, Stark shifts, Doppler shift, pressure shift, and black-body radiation shift. The measurements yielded transition frequencies of \(\nu_s = {54611.16(1.00)^{stat}  (30)^{sys}} \,\,{GHz}\) and \(\nu_t = {49881.35 (57)^{stat} (30)^{sys}} \,\,{GHz}\). Through these precise measurements and subsequent analysis, the researchers determined the proton charge radius to be \(r_E = \SI{0.84087 \pm 0.00039}{fm}\), achieving an unprecedented level of precision that represents a significant advancement in our understanding of proton structure and quantum electrodynamics.

 These findings underscore either unidentified systematic errors in previous measurements or the possibility of new physics beyond the Standard Model, laying the foundation for subsequent high-precision experimental efforts to resolve this fundamental inconsistency. Experimental Set-up is shown in Figure 6.

\begin{figure}[htbp]
    \centering
    \includegraphics[width=0.8\columnwidth]{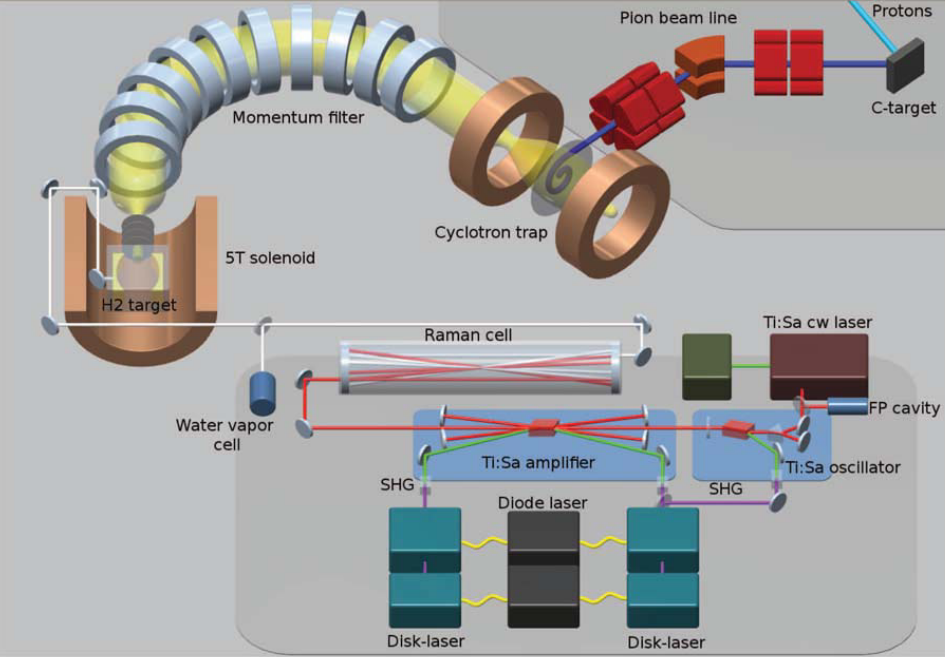}
    \caption{Negative pions from the accelerator are transported to a cyclotron trap, where they decay into MeV-energy muons. These muons are decelerated by a thin foil and guided by a 0.15-T toroidal magnetic field to a 5-T solenoid containing the hydrogen target. A muon entrance detector triggers a laser system that irradiates the formed muonic hydrogen (\(\mu p\)) to induce the 2S–2P transition within 0.9 ms. The laser system consists of Q-switched disk lasers operating in prelasing mode, frequency-doubled via second harmonic generation (SHG), and pumping a Ti:Sa laser. The Ti:Sa oscillator emits \(\sim\)700-nm pulses, converted to 5.5–6 \(\mu\)m via Raman shifts in hydrogen gas. Calibration from 5.5 to 6 \(\mu\)m was achieved using water vapor spectroscopy \cite{antognini2013proton}}
    \label{fig:proton_radius}
\end{figure}

Together, these studies provide critical insights into the proton's structure and highlight ongoing challenges in reconciling measurement techniques.

\subsection{Electron-Proton (ep) Scattering}
Electron-proton scattering is a technique used to probe the internal structure of protons by bombarding them with high-energy electrons. When electrons collide with protons, they can scatter off, providing information about the proton's charge distribution and other properties. This scattering process helps researchers determine fundamental aspects of protons, such as their form factors and radius \cite{xiong2019small}.

In the study conducted by Bernauer \textit{et al.} \cite{bernauer2010high}, the elastic electron-proton scattering cross-sections were measured at the Mainz Microtron (MAMI) under highly controlled conditions. Beam energies ranging from 180 to 855 MeV were employed, covering a momentum transfer squared range of $Q^2 = 0.004 \, \text{to} \, 1 \, (\text{GeV}/c)^2$, and approximately 1,400 cross-sections were recorded with statistical uncertainties below 0.2\%. To ensure reliability, high redundancy was implemented through repeated measurements of scattering angles using three high-resolution spectrometers. The data were analyzed using the Rosenbluth formula, which relates the differential cross-section to the electric ($G_E$) and magnetic ($G_M$) form factors through a combination of kinematic parameters:
\begin{equation}
    \left( \frac{d\sigma}{d\Omega} \right) = \left( \frac{d\sigma}{d\Omega} \right)_{\text{Mott}} \frac{\epsilon G_E^2 + \tau G_M^2}{\epsilon (1 + \tau)}
\end{equation}

Here, $G_E$ and $G_M$ are the electric and magnetic Sachs form factors, respectively, $m_p$ is the mass of the proton, $\tau = Q^2/(4m_p^2c^2)$, and $\epsilon = \left( 1 + 2 (1 + \tau) \tan^2(\theta/2) \right)^{-1}$, with $\theta$ being the angle of scattering of electrons. The least squares fitting was performed using the models: Single Dipole, Double Dipole, Simple Polynomial, Polynomial × Dipole, Polynomial + Dipole, Inverse Polynomial, Plain Uniform Cubic Spline, Cubic Spline × Dipole, Friedrich-Walcher Parametrization, and Extended Gari-Krümpelmann Model. The details of these models can be seen in \cite{bernauer2010measurement}.

And the charge radius was determined to be:
\[
\sqrt{\langle r_E^2 \rangle} = 0.879(5)_{\text{stat}}(4)_{\text{syst}}(2)_{\text{model}}(4)_{\text{group}} \, \text{fm}
\]

This value exceeds the standard dipole, but aligns well with the Coulomb-distortion-corrected Mainz result (0.880 $\pm$ 0.015 fm \cite{rosenfelder2000coulomb}) and Lamb shift experiments (0.883 $\pm$ 0.014 fm \cite{melnikov2000three}, 0.890 $\pm$ 0.014 fm \cite{udem1997phase, karshenboim1999we}). This finding had provided critical insights into the nucleon’s electromagnetic structure and the accuracy of various form factor models.

Further contributing to this understanding, a high-precision measurement of the proton elastic form factor ratio, $\mu_p \frac{G_E}{G_M}$, was conducted by  Zhan \textit{et al.} \cite{zhan2011high} at Jefferson Lab (JLab) Hall A, utilizing a 1.2 GeV polarized electron beam with currents ranging from 4 to 15 $\mu$A. The experiment employed advanced recoil polarimetry to determine the form factor ratio across four-momentum transfer squared values $Q^2 = 0.30 - 0.70 \, (\text{GeV}/c)^2$, with a total uncertainty of approximately 1\%. Scattering events were recorded using the BigBite spectrometer and the left High-Resolution Spectrometer (LHRS), while the recoil proton polarization was measured by a focal plane polarimeter (FPP). The form factor ratio was derived using the polarization transfer technique, expressed by the equation:

\begin{equation}
    \mu_p \frac{G_E}{G_M} = -\mu_p \frac{E_e + E_e'}{2M_p} \tan\left(\frac{\theta_e}{2}\right) \frac{P_t}{P_l}
\end{equation}

where $P_t$ and $P_l$ represent the transverse and longitudinal components of the proton polarization, respectively. The study revealed that the previously observed deviation of the ratio from unity at high $Q^2$ extends to the lowest $Q^2$ measured, offering new insights into the behavior of the form factor at low momentum transfers. The data analysis involved applying cuts to select clean elastic events and minimize systematic uncertainties, particularly in spin precession modeling. No corrections for two-photon exchange (TPE) effects were applied, as they were found to be negligible within the studied $Q^2$ range.
Key results from the study include an updated global fit that incorporates the measured data, revealing that the electric form factor is approximately 2\% smaller than earlier global fits. Additionally, the extracted proton charge radius was $\sqrt{\langle r^2_E \rangle} = 0.875 \pm 0.010 \, \text{fm}$, aligning with recent electron-proton interaction studies and atomic hydrogen Lamb shift measurements. These findings also suggest a continued role for the pion cloud at low $Q^2$, providing further understanding of the proton’s internal charge distribution. The results not only support the extension of the high-$Q^2$ observations but also contribute to refining the global understanding of proton structure at low $Q^2$. \\

Building on these insights, a recent study conducted by Xiong \textit{et al.} \cite{xiong2019small}, a high-precision electron-proton scattering experiment at the Jefferson Lab to determine the proton's charge radius, indicated a smaller proton radius of about 0.831 fm, significantly lower than the previously recommended value of around 0.88 fm \cite{mohr2012codata}. By using a novel technique with improved control over systematic uncertainties, they provided strong evidence supporting the smaller radius initially suggested by muonic hydrogen spectroscopy. For the data analysis used in this study, the reduced cross-section was calculated using the formula:

\begin{equation}
    \sigma_{\text{reduced}} = \frac{d\sigma}{d\Omega}_{e-p} \Bigg/ \left[ \frac{d\sigma}{d\Omega}_{\text{point-like}} \left( \frac{4M_p^2 E'/E}{4M_p^2 + Q^2} \right) \right]
\end{equation}

Here, $E$ and $E'$ are the energies of the incident and scattered electron, respectively, $M_p$ is the proton mass, $\Omega$ is the solid angle of the detector, and $Q^2$ is the momentum transfer. Dividing the kinematic factor yields $\sigma_{\text{reduced}}$, a linear combination of squared electromagnetic form factors.

And for the extraction of proton electric form factor ($G_E^p$), they used the Rational (1,1) function:

\begin{equation}
    f(Q^2) = nG_E(Q^2) = n \left( \frac{1 + p_1 Q^2}{1 + p_2 Q^2} \right)
\end{equation}

where, $n$ is the floating normalization parameter, $p_1$ and $p_2$ are the fit parameters.

and the proton charge radius is calculated using the equation,

\begin{equation}
    r_p = \sqrt{6(p_2 - p_1)}
\end{equation}

This analysis yielded $r_p = 0.831 \pm 0.007_{\text{stat}} \pm 0.012_{\text{syst}}$ femtometers, with a total systematic relative uncertainty of 1.4\%. The experiment's robustness was validated through comprehensive Monte Carlo simulations using the Geant4 toolkit \cite{collaboration2003geant4}, incorporating both e$^-$p and e$^-$e$^-$ event generators and including inelastic e$^-$p scattering events. Their findings have reignited discussions on the accuracy of existing measurement techniques and the true size of the proton. The experimental set-up is shown in Figure 7.

\begin{figure}[htbp]
    \centering
    \includegraphics[width=0.8\columnwidth]{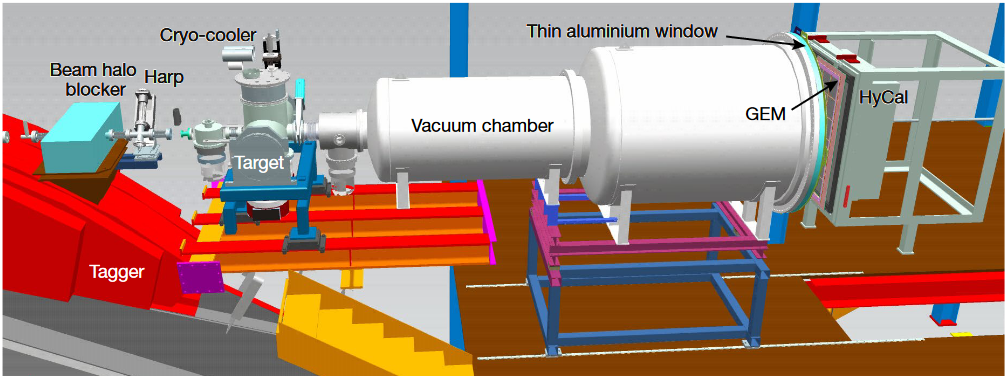} 
    \caption{A 4-cm-long windowless, cryo-cooled hydrogen gas flow target with an areal density of \( 2 \times 10^{18} \, \text{atoms/cm}^2 \) eliminates beam background from target windows. This setup is complemented by the HyCal20 hybrid electromagnetic calorimeter, which offers high resolution, large acceptance, and complete azimuthal coverage for forward-scattering angles, enabling simultaneous detection of electron pairs from \( e^- e^- \) scattering. Positioned in front of HyCal are two high-resolution X–Y gas electron multiplier (GEM) coordinate detectors. The entire system is enclosed within a two-section vacuum chamber spanning 5.5 m from the target to the detectors \cite{xiong2019small}.}
    \label{fig:image_label}
\end{figure}

In a related exploration of experimental precision, a study reported by Mihovilovic and Merkel \cite{mihovilovivc2019isr} of ISR Experiment at A1-Collaboration, initial experimental measurements were refined through analysis of cryogenic depositions on liquid hydrogen target walls, revealing side-wall layers approximately 200 times thicker (\(\sim0.1 \, \mu\text{m}\)) than front/back walls. This correction achieved sub-percent agreement between data and simulation across a \SI{200}{\mega\electronvolt} elastic line range, validating both radiative corrections and the Initial State Radiation (ISR) approach for low \(Q^2\) form-factor measurements. New data was obtained in the previously unexplored region of \(1 \cdot 10^{-3} \leq Q^2 \leq 4 \cdot 10^{-3} \, (\text{GeV}/c)^2\), yielding a proton radius of \(r_p = 0.836 \pm 0.017_{\text{stat}} \pm 0.059_{\text{syst}} \pm 0.003_{\Delta a, \Delta b} \, \text{fm}\). While current systematic uncertainties limit this measurement's contribution to the proton radius puzzle, the validated methodology establishes a foundation for future refined experiments \cite{mihovilovivc2019isr}.

\subsection{Reanalysis of Existing Data}

Building upon the previous methods, the reanalysis of existing experimental data plays a crucial role in refining our understanding of the proton radius. By revisiting earlier measurements with improved techniques and more precise models, discrepancies can be addressed, enhancing the accuracy of results.

Griffioen \textit{et al.} \cite{griffioen2015electron} employed polynomial expansions of the form factor, converging effectively below squared momentum transfers of \( 4m_\pi^2 \) (where \( m_\pi \) is the pion mass). By optimizing the number of terms to balance fit accuracy and overfitting, they obtained a proton radius of \( r_E = 0.840(16) \, \text{fm} \). This value aligns with muonic atom results but is smaller than CODATA. Building on this foundation, Hill \textit{et al.} \cite{hill2010model} employed a model-independent approach that integrated analyticity constraints with experimental electron-proton scattering data. Their analysis, which considered electron-neutron and pion-pion data, resulted in \( r_E = 0.870 \pm 0.023 \pm 0.012 \, \text{fm} \) for proton-only data, which agrees with that of Bernauer \textit{et al.} \cite{bernauer2010high}. However, in another reanalysis of electron scattering data from Mainz, Jefferson Lab, Saskatoon, and Stanford, Higinbotham \textit{et al.} \cite{higinbotham2016proton} employed stepwise regression and multivariate error estimation. They found that both low-momentum (\( Q^2 \)) data and a Padé extrapolation technique yielded a proton radius consistent with the muonic hydrogen result of \( 0.84 \, \text{fm} \). This analysis suggests that atomic hydrogen measurements are the outliers in the proton radius puzzle. Moreover, another study applied a fit function grounded in conformal mapping \cite{lorenz2014reduction}, they modeled the scattering data with high precision, obtaining a radius value consistent with muonic hydrogen measurements \cite{pohl2010size}. 

Later, Mihovilovic \textit{et al.} \cite{mihovilovivc2020reinterpretation} revisited the 1963 proton radius measurement of \( 0.805(11) \, \text{fm} \) \cite{hand1963electric} and identified an error in the original data analysis. Using modern computational techniques and a Padé (0,1) approximant for extraction, they updated the proton radius to \( r_p = 0.851(19) \, \text{fm} \), aligning with high-precision measurements from the  hydrogen spectroscopy. In a more comprehensive approach, Alarcon \textit{et al.} \cite{alarcon2019proton} combined chiral effective field theory with dispersion analysis. This framework linked the form factor’s finite-\( Q^2 \) behavior with its derivative at \( Q^2 = 0 \) through complex analyticity, avoiding challenges of \( Q^2 \to 0 \) extrapolation. They reported a radius of \( 0.844(7) \, \text{fm},\) consistent with muonic hydrogen measurements. Further refining this approach, Alarcon \textit{et al.} \cite{alarcon2020precise} extended their analysis to high-precision electron-proton elastic scattering data using a theoretical framework combining dispersion analysis and chiral effective field theory, constraining results up to \( Q^2 \approx 0.5 \, \text{GeV}^2 \). This yielded \( r_p^E = 0.842 \pm 0.02 \) (fit uncertainty) \((1\sigma)_{-0.002}^{+0.005} \, \text{fm},\) with reduced dependence on empirical fits.

Another study by Arrington \textit{et al.} \cite{arrington2007global} provided a critical evaluation of proton radius extractions, highlighting methodological inconsistencies and dataset tensions.They pointed out that many proton radius extractions have overlooked detailed assessments of model dependence and uncertainties. As a result, a wide range of proton radius values has been reported, often with underestimated uncertainties, further complicating efforts to resolve the proton radius puzzle. They reassessed charge radius extractions from global examination of elastic e-p scattering data and ultimately recommending an rms proton charge radius of \( r_p = 0.879 \pm 0.011 \, \text{fm} \), incorporating uncertainties that reflect dataset tensions and methodological inconsistencies. In a comparable analysis, Atac \textit{et al.} \cite{atac2021charge} conducted a global analysis incorporating flavor decomposition under charge symmetry. They derived a proton radius of \( \langle r_p \rangle = 0.852 \pm 0.002 \, \text{fm} \, (\text{stat.}) \pm 0.009 \, \text{fm} \, (\text{syst.}), \) consistent with muonic hydrogen spectroscopy.

The ongoing discrepancies in the proton radius puzzle reflect the need for more precise methodologies to resolve these conflicting results from various experiments. In response to these challenges, Lin \textit{et al.} \cite{lin2022new} employed a constrained Gaussian process in a combined analysis of space- and timelike regions. This approach, which utilized dispersion theory, yielded \( r_p^E = 0.840_{-0.002}^{+0.003} \, \text{fm},\) aligning with Lamb shift and hyperfine splitting measurements. Similarly, Gramolin \textit{et al.} \cite{gramolin2022transverse} introduced a novel method relating the proton radius to its transverse charge density, applying it to A1 Collaboration’s data. Their analysis yielded a proton radius of \( r_E = 0.889 \pm 0.005_{\text{stat}} \pm 0.005_{\text{syst}} \pm 0.004_{\text{model}} \, \text{fm} \), which was consistent with the original result. The reanalysis showed that the proton radius discrepancy could not be explained by issues with fitting or extrapolating the A1 data to \( Q^2 = 0 \). However, Boone \textit{et al.} \cite{boone2023comment} pointed out that applying the extraction procedure from Gramolin and Russell \cite{gramolin2022transverse} to different datasets produced inconsistent results. In particular, re-binning the same data led to vastly different radii. Despite exploiting the known \( Q^2 \) limiting behavior, the method failed to extrapolate beyond the data included in the fits. Boone \textit{et al.} concluded that this method was not robust and that the strong claims made regarding the proton radius were not justified.

The persistent inconsistencies in the proton radius puzzle highlight the necessity for more accurate methods to reconcile the conflicting findings from different experiments. For that reason, Zhou \textit{et al.} \cite{zhou2019reexamining} introduced a new nonparametric approach using a constrained Gaussian process to estimate the proton charge radius from electron scattering data. This Bayesian method models the electric form factor without predefined assumptions, applying two physical constraints: normalization at zero momentum transfer and a monotonically decreasing form factor shape. The analysis yielded a radius of \( r_p = 0.845 \pm 0.001 \, \text{fm} \), consistent with results from muonic hydrogen. However, the study also revealed that the radius value depends significantly on the specific constraints and data range applied. Building on the variability in proton radius measurements, Horbatsch \textit{et al.} \cite{horbatsch2016evaluation} conducted a reanalysis of electron-proton scattering data \cite{bernauer2010high, arrington2007global}, using low-\( Q^2 \) cross-sections with two single-parameter models: one with dipole parametrization and the other employing conformal mapping. This analysis yielded proton charge radius values of 0.84 and 0.89 fm. Extending the fits to all \( Q^2 \) with cubic splines confirmed this range. However, while these values are consistent with some measurements, they fail to resolve the discrepancy observed with results from muonic and electronic hydrogen spectroscopy, further complicating efforts to reconcile different experimental methods.

\section{Results and Discussion}

The extensive review of experimental measurements and theoretical analyses reveals a complex landscape in the determination of the proton radius, with significant implications for our understanding of fundamental physics. The historical consensus of approximately 0.88 fm \cite{mohr1999codata}, derived from electron-proton scattering and traditional hydrogen spectroscopy, was dramatically challenged by the 2010 muonic hydrogen measurements \cite{pohl2010size} that yielded a significantly smaller radius of about 0.84 fm. This discrepancy, initially viewed as potentially revolutionary for quantum electrodynamics (QED), has sparked a decade of intensive research and methodological refinement. Recent high-precision measurements, particularly the work of Xiong \textit{et al.} \cite{xiong2019small}, using electron-proton scattering at Jefferson Lab, have produced results (0.831 fm) that align closely with the muonic hydrogen value, suggesting that earlier electron-based measurements may have contained unrecognized systematic errors.

The reanalysis of existing data has played a crucial role in understanding and potentially resolving the proton radius puzzle. Notable studies by Zhou \textit{et al.} \cite{zhou2019reexamining}, employing a novel nonparametric approach using constrained Gaussian processes, yielded a radius of \(0.845 \pm 0.001 \, \text{fm}\), while Alarcon \textit{et al.}'s \cite{alarcon2019proton} theoretical framework combining dispersion analysis and chiral effective field theory produced a value of \(0.842 \pm 0.02 \, \text{fm}\). These results, along with other recent analyses, demonstrate a growing convergence toward the smaller radius value initially suggested by muonic hydrogen spectroscopy \cite{pohl2010size}. This convergence is further supported by Atac \textit{et al.}'s \cite{atac2021charge} global analysis of elastic form factors, which determined a radius of \(0.852 \pm 0.002 \, \text{(stat.)} \pm 0.009 \, \text{(syst.)} \, \text{fm}\), indicating that careful consideration of systematic effects and improved analytical methods can reconcile apparently conflicting measurements.

Recent advances in experimental techniques have significantly enhanced our ability to probe proton structure with unprecedented precision. The application of two-photon ultraviolet direct frequency comb spectroscopy by Grinin \textit{et al.} \cite{grinin2020two}. on the 1S-3S transition in atomic hydrogen yielded a proton charge radius of \(0.8482(38) \, \text{fm}\), while Bezginov \textit{et al.}'s \cite{bezginov2019measurement} direct measurement of the \(n = 2\) Lamb shift in atomic hydrogen produced a value of \(0.833 \pm 0.010 \, \text{fm}\). These results, obtained through different experimental approaches, demonstrate remarkable consistency with the muonic hydrogen measurements and suggest that the original "puzzle" may have stemmed from underestimated systematic uncertainties in earlier experiments rather than indicating new physics beyond the Standard Model.

The collective evidence from multiple experimental techniques and theoretical analyses increasingly supports a proton radius value closer to \(0.84 \, \text{fm}\), challenging the previously accepted larger value of \(0.88 \, \text{fm}\). However, some tension remains, as demonstrated by Gramolin's \cite{gramolin2022transverse} analysis yielding a radius of \(0.889(5)_{\text{stat.}} (5)_{\text{syst.}} (4)_{\text{model}} \, \text{fm}\) from A1 Collaboration data. This persistent disagreement, though diminished, highlights the importance of continuing refinements in both experimental techniques and theoretical frameworks. The resolution of the proton radius puzzle appears to lie not in new physics beyond the Standard Model, as initially speculated, but rather in our improved understanding of systematic effects and the development of more sophisticated analytical methods. This conclusion has significant implications for our confidence in QED and the Standard Model, while simultaneously demonstrating the importance of rigorous experimental and analytical approaches in modern physics.

\begin{figure}[htbp]
    \centering
    \includegraphics[width=0.8\columnwidth]{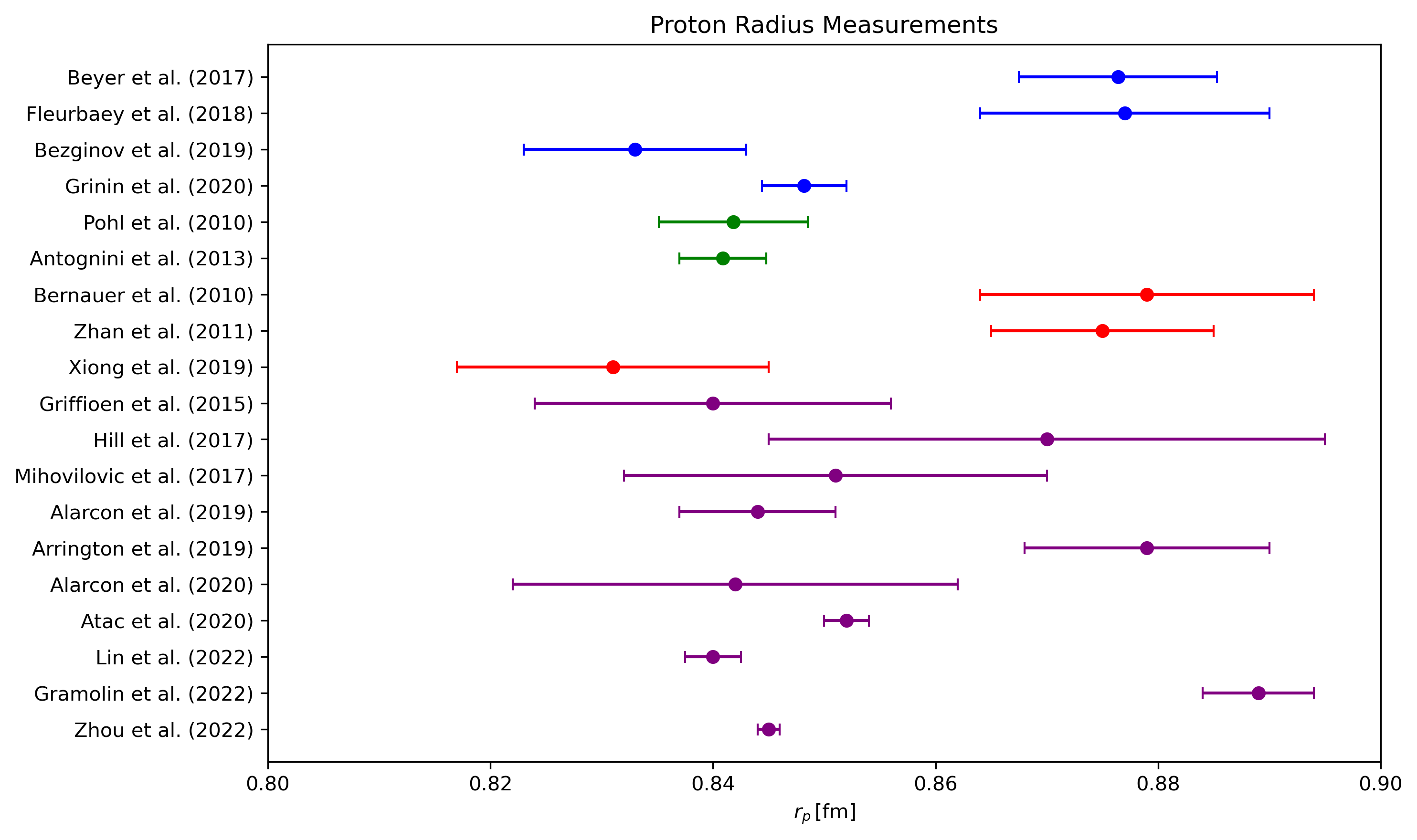}
    \caption{Proton radius measurements categorized by technique. The blue points represent results from Hydrogen Spectroscopy, the green points are from $\mu$-Hydrogen Spectroscopy, the red points are from Ep Scattering, and the purple points are from Reanalysis of data.}
    \label{fig:proton_radius}
\end{figure}

Building on this improved understanding, the field of nuclear and particle physics continues to pursue groundbreaking experimental and theoretical initiatives to advance our knowledge of atomic structure and fundamental constants. Three major collaborations—CREMA\cite{amaro2022laser}, FAMU \cite{pizzolotto2020famu, pizzolotto2021measurement}, and J-PARC/Riken \cite{schmidt2014proceedings}—are working towards achieving unprecedented precision in measuring the $1S$ hyperfine splitting (HFS) in muonic hydrogen, while an upgraded CREMA 2010 setup \cite{pohl2010size, antognini2013proton} promises to enhance $2S-2P$ measurements significantly. Meanwhile, two research groups \cite{dreissen2020ramsey, herrmann2009feasibility} are investigating He$^+$ $1S-2S$ transitions using cutting-edge frequency comb and trap technologies, which could help resolve the proton radius puzzle and test quantum electrodynamics (QED) contributions. Complementing these efforts, various scattering experiments, including the upgraded PRad experiment (PRad-II), MUSE (PSI), AMBER (CERN), and PRES, are being conducted to measure the proton radius with improved accuracy \cite{antognini2013proton}. On the theoretical front, significant progress is anticipated in several areas: lattice Quantum Chromodynamics (QCD) calculations are approaching experimental precision levels, $\chi$PT calculations are being extended to muonic atoms, and efforts are underway to achieve parts per million (ppm) accuracy in $\mu$H HFS predictions \cite{antognini2013proton}. Furthermore, improvements in QED calculations for hydrogen S-levels and refined understanding of nucleon $2\gamma$-exchange contributions, particularly for neutrons, will enhance our ability to test theoretical predictions and extract fundamental constants with unprecedented precision.

\section{Conclusion}

The proton radius puzzle remains a central challenge in modern physics, reflecting both the precision and limitations of current experimental and theoretical methods. Recent studies, such as those by Xiong \textit{et al.} \cite{xiong2019small} and Bezginov \textit{et al.} \cite{bezginov2019measurement}, converge on a smaller proton radius of approximately 0.84 femtometers \cite{pohl2010size}, consistent with muonic hydrogen results, yet discrepancies with earlier electron-proton scattering and atomic hydrogen spectroscopy measurements persist, suggesting the presence of systematic errors or limitations in theoretical frameworks, as noted by Arrington and Sick \cite{arrington2007global}. The divergence between results from muonic and electronic hydrogen systems remains unexplained, with potential contributions from both experimental uncertainties and theoretical gaps. Works by Zhou \textit{et al.} \cite{zhou2019reexamining} and Grinin \textit{et al.} \cite{grinin2020two} emphasize the need for improved data analysis methods and theoretical modeling to resolve these discrepancies. 

The implications extend beyond the puzzle, potentially challenging QED and the Standard Model. Solving the proton radius puzzle requires a coordinated approach, with experimental efforts focused on reducing systematic uncertainties through advanced setups, such as improved scattering experiments at Jefferson Lab \cite{xiong2019small} and next-generation spectroscopy techniques \cite{beyer2017rydberg}. Concurrent theoretical advancements, including the work of Alarcón \textit{et al.} \cite{alarcon2019proton} and Lin \textit{et al.} \cite{lin2022new}, are essential for addressing these inconsistencies and exploring extensions to QED. By integrating experimental precision with theoretical rigor, the field can advance toward resolving this enduring puzzle, yielding deeper insights into the fundamental forces of nature.

\bibliographystyle{unsrt}

\bibliography{ref}

\begin{thebibliography}{10}

\bibitem{xiong2019small}
W~Xiong, A~Gasparian, H~Gao, D~Dutta, M~Khandaker, N~Liyanage, E~Pasyuk, C~Peng, X~Bai, L~Ye, et~al.
\newblock A small proton charge radius from an electron--proton scattering experiment.
\newblock {\em Nature}, 575(7781):147--150, 2019.

\bibitem{bezginov2019measurement}
N~Bezginov, T~Valdez, M~Horbatsch, A~Marsman, AC~Vutha, and EA~Hessels.
\newblock A measurement of the atomic hydrogen lamb shift and the proton charge radius.
\newblock {\em Science}, 365(6457):1007--1012, 2019.

\bibitem{pohl2010size}
Randolf Pohl, Aldo Antognini, Fran{\c{c}}ois Nez, Fernando~D Amaro, Fran{\c{c}}ois Biraben, Jo{\~a}o~MR Cardoso, Daniel~S Covita, Andreas Dax, Satish Dhawan, Luis~MP Fernandes, et~al.
\newblock The size of the proton.
\newblock {\em nature}, 466(7303):213--216, 2010.

\bibitem{arrington2015evaluation}
John Arrington and Ingo Sick.
\newblock Evaluation of the proton charge radius from electron--proton scattering.
\newblock {\em Journal of Physical and Chemical Reference Data}, 44(3), 2015.

\bibitem{alarcon2019proton}
Jose~Manuel Alarcon, Douglas~W Higinbotham, Christian Weiss, and Zhihong Ye.
\newblock Proton charge radius extraction from electron scattering data using dispersively improved chiral effective field theory.
\newblock {\em Physical Review C}, 99(4):044303, 2019.

\bibitem{lin2022new}
Yong-Hui Lin, Hans-Werner Hammer, and Ulf-G Mei{\ss}ner.
\newblock New insights into the nucleon’s electromagnetic structure.
\newblock {\em Physical Review Letters}, 128(5):052002, 2022.

\bibitem{pohl2013muonic}
Randolf Pohl, Ronald Gilman, Gerald~A Miller, and Krzysztof Pachucki.
\newblock Muonic hydrogen and the proton radius puzzle.
\newblock {\em Annual Review of Nuclear and Particle Science}, 63(1):175--204, 2013.

\bibitem{carlson2015proton}
Carl~E Carlson.
\newblock The proton radius puzzle.
\newblock {\em Progress in Particle and Nuclear Physics}, 82:59--77, 2015.

\bibitem{mohr1999codata}
Peter~J Mohr and Barry~N Taylor.
\newblock Codata recommended values of the fundamental physical constants: 1998.
\newblock {\em Journal of Physical and Chemical Reference Data}, 28(6):1713--1852, 1999.

\bibitem{sick2003rms}
Ingo Sick.
\newblock On the rms-radius of the proton.
\newblock {\em Physics Letters B}, 576(1-2):62--67, 2003.

\bibitem{bernauer2010high}
Jan~C Bernauer, P~Achenbach, C~Ayerbe~Gayoso, R~B{\"o}hm, Damir Bosnar, L~Debenjak, MO~Distler, L~Doria, A~Esser, H~Fonvieille, et~al.
\newblock High-precision determination of the electric and magnetic form factors of the proton.
\newblock {\em Physical Review Letters}, 105(24):242001, 2010.

\bibitem{beyer2017rydberg}
Axel Beyer, Lothar Maisenbacher, Arthur Matveev, Randolf Pohl, Ksenia Khabarova, Alexey Grinin, Tobias Lamour, Dylan~C Yost, Theodor~W H{\"a}nsch, Nikolai Kolachevsky, et~al.
\newblock The rydberg constant and proton size from atomic hydrogen.
\newblock {\em Science}, 358(6359):79--85, 2017.

\bibitem{rutherford2010collision}
Ernest Rutherford.
\newblock Collision of $\alpha$ particles with light atoms. iv. an anomalous effect in nitrogen.
\newblock {\em Philosophical Magazine}, 90(S1):31--37, 2010.

\bibitem{yang2024discovery}
HB~Yang, ZG~Gan, YJ~Li, ML~Liu, SY~Xu, C~Liu, MM~Zhang, ZY~Zhang, MH~Huang, CX~Yuan, et~al.
\newblock Discovery of new isotopes os 160 and w 156: Revealing enhanced stability of the n= 82 shell closure on the neutron-deficient side.
\newblock {\em Physical Review Letters}, 132(7):072502, 2024.

\bibitem{karr2019progress}
Jean-Philippe Karr and Dominique Marchand.
\newblock Progress on the proton-radius puzzle, 2019.

\bibitem{bernauer2020proton}
Jan~C Bernauer.
\newblock The proton radius puzzle--9 years later.
\newblock In {\em EPJ Web of Conferences}, volume 234, page 01001. EDP Sciences, 2020.

\bibitem{antognini2013proton}
Aldo Antognini, Fran{\c{c}}ois Nez, Karsten Schuhmann, Fernando~D Amaro, Francois Biraben, Jo{\~a}o~MR Cardoso, Daniel~S Covita, Andreas Dax, Satish Dhawan, Marc Diepold, et~al.
\newblock Proton structure from the measurement of 2s-2p transition frequencies of muonic hydrogen.
\newblock {\em Science}, 339(6118):417--420, 2013.

\bibitem{hofstadter1956electron}
Robert Hofstadter.
\newblock Electron scattering and nuclear structure.
\newblock {\em Reviews of Modern Physics}, 28(3):214, 1956.

\bibitem{bernauer2014electric}
JC~Bernauer, MO~Distler, J~Friedrich, Th~Walcher, P~Achenbach, C~Ayerbe~Gayoso, R~B{\"o}hm, Damir Bosnar, L~Debenjak, L~Doria, et~al.
\newblock Electric and magnetic form factors of the proton.
\newblock {\em Physical Review C}, 90(1):015206, 2014.

\bibitem{zhan2011high}
Xiaohui Zhan, K~Allada, DS~Armstrong, J~Arrington, W~Bertozzi, W~Boeglin, J-P Chen, K~Chirapatpimol, S~Choi, E~Chudakov, et~al.
\newblock High-precision measurement of the proton elastic form factor ratio $\mu$pge/gm at low q2.
\newblock {\em Physics Letters B}, 705(1-2):59--64, 2011.

\bibitem{biraben2009spectroscopy}
Fran{\c{c}}ois Biraben.
\newblock Spectroscopy of atomic hydrogen: How is the rydberg constant determined?
\newblock {\em The European Physical Journal Special Topics}, 172(1):109--119, 2009.

\bibitem{bransden2003physics}
Brian~Harold Bransden and Charles~Jean Joachain.
\newblock {\em Physics of atoms and molecules}.
\newblock Pearson Education India, 2003.

\bibitem{woodgate1970elementary}
Gordon~Kemble Woodgate.
\newblock Elementary atomic structure.
\newblock 1970.

\bibitem{baklanov1974precise}
EV~Baklanov and VP~Chebotaev.
\newblock On the precise measurement of the frequency transition 1s--2s of the hydrogen atom.
\newblock {\em Optics Communications}, 12(3):312--314, 1974.

\bibitem{fleurbaey2018new}
H{\'e}l{\`e}ne Fleurbaey, Sandrine Galtier, Simon Thomas, Marie Bonnaud, Lucile Julien, Fran{\c{c}}ois Biraben, Fran{\c{c}}ois Nez, Michel Abgrall, and Jocelyne Gu{\'e}na.
\newblock New measurement of the 1 s-3 s transition frequency of hydrogen: contribution to the proton charge radius puzzle.
\newblock {\em Physical review letters}, 120(18):183001, 2018.

\bibitem{thomas2018high}
Simon Thomas, H{\'e}l{\`e}ne Fleurbaey, Sandrine Galtier, Marie Bonnaud, Lucile Julien, Fran{\c{c}}ois Biraben, and Fran{\c{c}}ois Nez.
\newblock High resolution spectroscopy of 1s-3s transition in hydrogen with a cw laser.
\newblock In {\em 2018 Conference on Precision Electromagnetic Measurements (CPEM 2018)}, pages 1--2. IEEE, 2018.

\bibitem{mohr2005codata}
Peter~J Mohr and Barry~N Taylor.
\newblock Codata recommended values of the fundamental physical constants: 2002.
\newblock {\em Reviews of modern physics}, 77(1):1--107, 2005.

\bibitem{grinin2020two}
Alexey Grinin, Arthur Matveev, Dylan~C Yost, Lothar Maisenbacher, Vitaly Wirthl, Randolf Pohl, Theodor~W H{\"a}nsch, and Thomas Udem.
\newblock Two-photon frequency comb spectroscopy of atomic hydrogen.
\newblock {\em Science}, 370(6520):1061--1066, 2020.

\bibitem{mohr2008codata}
Peter~J Mohr, Barry~N Taylor, and David~B Newell.
\newblock Codata recommended values of the fundamental physical constants: 2006.
\newblock {\em Journal of Physical and Chemical Reference Data}, 37(3):1187--1284, 2008.

\bibitem{bernauer2010measurement}
Jan~C Bernauer.
\newblock Measurement of the elastic electron-proton cross section and separation of the electric and magnetic form factor in the q $\{$sup 2$\}$ range from 0.004 to 1 (gev/c)$\{$sup 2$\}$.
\newblock 2010.

\bibitem{rosenfelder2000coulomb}
R~Rosenfelder.
\newblock Coulomb corrections to elastic electron--proton scattering and the proton charge radius.
\newblock {\em Physics Letters B}, 479(4):381--386, 2000.

\bibitem{melnikov2000three}
Kirill Melnikov and Timo van Ritbergen.
\newblock Three-loop slope of the dirac form factor and the 1 s lamb shift in hydrogen.
\newblock {\em Physical review letters}, 84(8):1673, 2000.

\bibitem{udem1997phase}
Th~Udem, A~Huber, B~Gross, J~Reichert, M~Prevedelli, M~Weitz, and Th~W H{\"a}nsch.
\newblock Phase-coherent measurement of the hydrogen 1 s- 2 s transition frequency with an optical frequency interval divider chain.
\newblock {\em Physical review letters}, 79(14):2646, 1997.

\bibitem{karshenboim1999we}
Savely~G Karshenboim.
\newblock What do we actually know about the proton radius?
\newblock {\em Canadian Journal of Physics}, 77(4):241--266, 1999.

\bibitem{mohr2012codata}
Peter~J Mohr, Barry~N Taylor, and David~B Newell.
\newblock Codata recommended values of the fundamental physical constants: 2010.
\newblock {\em Journal of Physical and Chemical Reference Data}, 41(4), 2012.

\bibitem{collaboration2003geant4}
GEANT Collaboration, S~Agostinelli, et~al.
\newblock Geant4--a simulation toolkit.
\newblock {\em Nucl. Instrum. Meth. A}, 506(25):0, 2003.

\bibitem{mihovilovivc2019isr}
Miha Mihovilovi{\v{c}} and Harald Merkel.
\newblock Isr experiment at a1-collaboration.
\newblock In {\em EPJ Web of Conferences}, volume 218, page 04001. EDP Sciences, 2019.

\bibitem{griffioen2015electron}
Keith Griffioen, Carl Carlson, and Sarah Maddox.
\newblock Are electron scattering data consistent with a small proton radius?
\newblock {\em arXiv preprint arXiv:1509.06676}, 2015.

\bibitem{hill2010model}
Richard~J Hill and Gil Paz.
\newblock Model-independent extraction of the proton charge radius from electron scattering.
\newblock {\em Physical Review D—Particles, Fields, Gravitation, and Cosmology}, 82(11):113005, 2010.

\bibitem{higinbotham2016proton}
Douglas~W Higinbotham, Al~Amin Kabir, Vincent Lin, David Meekins, Blaine Norum, and Brad Sawatzky.
\newblock Proton radius from electron scattering data.
\newblock {\em Physical Review C}, 93(5):055207, 2016.

\bibitem{lorenz2014reduction}
IT~Lorenz and Ulf-G Mei{\ss}ner.
\newblock Reduction of the proton radius discrepancy by 3$\sigma$.
\newblock {\em Physics Letters B}, 737:57--59, 2014.

\bibitem{mihovilovivc2020reinterpretation}
Miha Mihovilovi{\v{c}}, Douglas~W Higinbotham, Melisa Bevc, and Simon {\v{S}}irca.
\newblock Reinterpretation of classic proton charge form factor measurements.
\newblock {\em Frontiers in Physics}, 8:36, 2020.

\bibitem{hand1963electric}
L~No Hand, D~Go Miller, and Richard Wilson.
\newblock Electric and magnetic form factors of the nucleon.
\newblock {\em Reviews of Modern Physics}, 35(2):335, 1963.

\bibitem{alarcon2020precise}
Jose~Manuel Alarc{\'o}n, DW~Higinbotham, and Christian Weiss.
\newblock Precise determination of the proton magnetic radius from electron scattering data.
\newblock {\em Physical Review C}, 102(3):035203, 2020.

\bibitem{arrington2007global}
J~Arrington, W~Melnitchouk, and JA~Tjon.
\newblock Global analysis of proton elastic form factor data with two-photon exchange corrections.
\newblock {\em Physical Review C—Nuclear Physics}, 76(3):035205, 2007.

\bibitem{atac2021charge}
H~Atac, M~Constantinou, Z-E Meziani, M~Paolone, and N~Sparveris.
\newblock Charge radii of the nucleon from its flavor dependent dirac form factors.
\newblock {\em The European Physical Journal A}, 57:1--8, 2021.

\bibitem{gramolin2022transverse}
Alexander~V Gramolin and Rebecca~L Russell.
\newblock Transverse charge density and the radius of the proton.
\newblock {\em Physical Review D}, 105(5):054004, 2022.

\bibitem{boone2023comment}
Benjamin Boone, Michael Chen, Kevin Sturm, Justin Yoo, and Douglas Higinbotham.
\newblock Comment on transverse charge density and the radius of the proton.
\newblock {\em arXiv preprint arXiv:2302.07356}, 2023.

\bibitem{zhou2019reexamining}
Shuang Zhou, P~Giulani, J~Piekarewicz, Anirban Bhattacharya, and Debdeep Pati.
\newblock Reexamining the proton-radius problem using constrained gaussian processes.
\newblock {\em Physical Review C}, 99(5):055202, 2019.

\bibitem{horbatsch2016evaluation}
M~Horbatsch and EA~Hessels.
\newblock Evaluation of the strength of electron-proton scattering data for determining the proton charge radius.
\newblock {\em Physical Review C}, 93(1):015204, 2016.

\bibitem{amaro2022laser}
Pedro Amaro, Andrzej Adamczak, Marwan Abdou~Ahmed, L~Affolter, Fernando~Domingues Amaro, Patricia Carvalho, T-L Chen, Lu{\'\i}s~MP Fernandes, M~Ferro, Damian Goeldi, et~al.
\newblock Laser excitation of the 1s-hyperfine transition in muonic hydrogen.
\newblock {\em SciPost Physics}, 13(2):020, 2022.

\bibitem{pizzolotto2020famu}
C~Pizzolotto, A~Adamczak, D~Bakalov, G~Baldazzi, M~Baruzzo, R~Benocci, R~Bertoni, M~Bonesini, V~Bonvicini, H~Cabrera, et~al.
\newblock The famu experiment: muonic hydrogen high precision spectroscopy studies.
\newblock {\em The European Physical Journal A}, 56:1--15, 2020.

\bibitem{pizzolotto2021measurement}
C~Pizzolotto, A~Sbrizzi, A~Adamczak, D~Bakalov, G~Baldazzi, M~Baruzzo, R~Benocci, R~Bertoni, M~Bonesini, H~Cabrera, et~al.
\newblock Measurement of the muon transfer rate from muonic hydrogen to oxygen in the range 70-336 k.
\newblock {\em Physics Letters A}, 403:127401, 2021.

\bibitem{schmidt2014proceedings}
Alexander Schmidt and Christian Sander.
\newblock Proceedings, 20th international conference on particles and nuclei (panic 14).
\newblock Technical report, LHC/ATLAS Experiment, 2014.

\bibitem{dreissen2020ramsey}
LS~Dreissen, C~Roth, EL~Gr{\"u}ndeman, JJ~Krauth, MGJ Favier, and KSE Eikema.
\newblock Ramsey-comb precision spectroscopy in xenon at vacuum ultraviolet wavelengths produced with high-order harmonic generation.
\newblock {\em Physical Review A}, 101(5):052509, 2020.

\bibitem{herrmann2009feasibility}
Maximilian Herrmann, M~Haas, Ulrich~D Jentschura, Franz Kottmann, Dietrich Leibfried, Guido Saathoff, Christoph Gohle, Akira Ozawa, Valentin Batteiger, Sebastian Kn{\"u}nz, et~al.
\newblock Feasibility of coherent xuv spectroscopy on the 1 s-2 s transition in singly ionized helium.
\newblock {\em Physical Review A—Atomic, Molecular, and Optical Physics}, 79(5):052505, 2009.

\end{thebibliography}

\end{document}